\documentclass[10pt,letterpaper]{article}
\usepackage[top=0.85in,left=2.75in,footskip=0.75in]{geometry}

\usepackage{amsmath,amssymb}

\usepackage{changepage}

\usepackage[utf8x]{inputenc}

\usepackage{textcomp,marvosym}

\usepackage{cite}

\usepackage{nameref,hyperref}

\usepackage{microtype}
\DisableLigatures[f]{encoding = *, family = * }

\usepackage[table]{xcolor}

\usepackage{array}

\newcolumntype{+}{!{\vrule width 2pt}}

\newlength\savedwidth

\raggedright
\usepackage[aboveskip=1pt,labelfont=bf,labelsep=period,justification=raggedright,singlelinecheck=off]{caption}

\makeatletter
\renewcommand{\@biblabel}[1]{\quad#1.}
\makeatother

\date{}

\usepackage{lastpage,fancyhdr,graphicx}
\usepackage{epstopdf}
\fancyheadoffset[L]{2.25in}
\fancyfootoffset[L]{2.25in}

\begin{document}
\vspace*{0.2in}

\begin{flushleft}
{\Large
\textbf\newline{Bayesian optimization for computationally extensive probability distributions} 
}
\newline
\\
Ryo Tamura\textsuperscript{1,2,3\P*},
Koji Hukushima\textsuperscript{2,4\P*}
\\
\bigskip
\textbf{1} International Center for Materials Nanoarchitectonics (WPI-MANA),
National Institute for Materials Science, 1-1 Namiki, Tsukuba, Ibaraki 305-0044, Japan
\\
\textbf{2} Research and Services Division of Materials Data and Integrated System,
National Institute for Materials Science, 1-2-1 Sengen, Tsukuba, Ibaraki 305-0047, Japan
\\
\textbf{3} Graduate School of Frontier Sciences, 
The University of Tokyo, 5-1-5 Kashiwanoha, Kashiwa, Chiba 277-8568, Japan
\\
\textbf{4} Department of Basic Science, Graduate School of Arts and Sciences, 
The University of Tokyo, Komaba, Meguro, Tokyo 153-8902, Japan
\\
\bigskip

%
%
$^\P$ These authors contributed equally to this work. \\






* Corresponding authors \\
E-mail: tamura.ryo@nims.go.jp \\
E-mail: hukusima@phys.c.u-tokyo.ac.jp

\end{flushleft}
\section*{Abstract}
An efficient method for finding a better maximizer of computationally extensive probability distributions is proposed on the basis of a Bayesian optimization technique.  
A key idea of the proposed method is to use extreme values of acquisition functions by Gaussian processes for the next training phase, 
which should be located near a local maximum or a global maximum of the probability distribution.
Our Bayesian optimization technique is applied to the posterior distribution in the effective physical model estimation,
which is a computationally extensive probability distribution. 
Even when the number of sampling points on the posterior distributions is fixed to be small,
the Bayesian optimization provides a better maximizer of the posterior distributions in comparison to those by the random search method, the steepest descent method, or the Monte Carlo method.
Furthermore,
the Bayesian optimization improves the results efficiently by combining the steepest descent method and thus it is a powerful tool to search for a better maximizer of computationally extensive probability distributions.

\section*{Introduction}

Bayesian optimization\cite{Mockus-1989,Jones-1998,Pelikan-1999,Snoek-2012,Ueno-2016} has recently attracted much attention as a method to search the maximizer/minimizer of a black-box function in informatics and materials science\cite{Seko-2013,Toyoura-2016,Kiyohara-2016,Balachandran-2016,Ju-2017,Packwood-2017,Seko-2017}.
In this method, the black-box function is interpolated by Gaussian processes.
Then the interpolated function is used to predict the maximizer/minimizer of the black-box function.
The Bayesian optimization is effective for problems where the value on the black-box function cannot be easily obtained. 
In other words, it is effective when the data for the black-box function is limited.


We are currently developing a generic effective physical model
estimation method from experimentally measured data using machine learning,
which relates to calibration in data science\cite{Kennedy-2001,Higdon-2004,Bayarri-2009}.
As the first example,
we developed a method to estimate a set of model parameters $\mathbf{x}
= (x_1, ..., x_K)$ in the Hamiltonian $\mathcal{H} (\mathbf{x})$, where
$K$ is the number of model parameters\cite{Tamura-2016}.
Let 
$\mathbf{y}^{\rm ex}$ be the set of physical quantities $\{ y^{\rm ex}(g_l) \}_{l=1, ..., L}$ depending on the external parameter $g_l$ with $L$ being the number of data.
By using Bayes' theorem, 
the posterior distribution $P(\mathbf{x}| \mathbf{y}^{\rm ex})$,
or the conditional probability of $\mathbf{x}$ given $\mathbf{y}^{\rm ex}$ is expressed as
\begin{eqnarray}
P(\mathbf{x}| \mathbf{y}^{\rm ex} )= \frac{P( \mathbf{y}^{\rm ex} |\mathbf{x}) P(\mathbf{x})}{Z(\mathbf{y}^{\rm ex})}, 
\label{eq:Bayes_theorem}
\end{eqnarray}
where $P(\mathbf{x})$ and $Z(\mathbf{y}^{\rm ex})$
are the prior distributions of the model parameters and the normalization constant of the posterior distribution, respectively.
Assuming that the observed noise follows a Gaussian distribution with a mean of zero and a standard deviation of $\sigma$, 
the likelihood function $P(\mathbf{y}^{\rm ex}|\mathbf{x})$ is given as 
\begin{eqnarray}
P(\mathbf{y}^{\rm ex} |\mathbf{x}) \propto \exp \left[ - \frac{1}{2 \sigma^2} \sum_{l=1}^L \left( y^{\rm ex}(g_l) - y^{\rm cal} (g_l, \mathbf{x}) \right)^2 \right], \label{eq:cond_phys_qu}
\end{eqnarray}
where $\{ y^{\rm cal} (g_l, \mathbf{x}) \}_{l=1, \cdots, L}$ is the $g_l$ dependence of the physical quantity calculated from $\mathcal{H} (\mathbf{x})$,
and hereinafter let $\mathbf{y}^{\rm cal} (\mathbf{x})$ be the set of $\{ y^{\rm cal} (g_l, \mathbf{x}) \}_{l=1, \cdots, L}$.
Then, 
the posterior distribution is expressed as
\begin{eqnarray}
P(\mathbf{x}|\mathbf{y}^{\rm ex}) \propto \exp \left[ -E(\mathbf{x})\right],  \label{eq:cond_model}
\end{eqnarray}
where the ``energy function'' as a function of $\mathbf{x}$ is given by
\begin{eqnarray}
E (\mathbf{x}) = \frac{1}{2 \sigma^2} \sum_{l=1}^L \left( y^{\rm ex}(g_l) - y^{\rm cal} (g_l, \mathbf{x}) \right)^2 - \log P (\mathbf{x}). \label{eq:energy_function}
\end{eqnarray}
From the viewpoint of the maximum a posterior (MAP) estimation,
the plausible model parameters for explaining $\mathbf{y}^{\rm ex}$ 
are obtained as the maximizer of Eq.~(\ref{eq:cond_model}) or the minimizer of Eq.~(\ref{eq:energy_function}).
Thus,
the most fundamental task for construction of an effective model is summarized to maximize Eq.~(\ref{eq:cond_model}) or minimize Eq.~(\ref{eq:energy_function}).


A computational method to evaluate the posterior distribution or energy function consists of a double-loop calculation. 
In the inner loop, the physical quantities $\mathbf{y}^{\rm cal} (\mathbf{x})$ are calculated from $\mathcal{H} (\mathbf{x})$ when a set of model parameters is given.
The computational cost of the inner loop depends on the simulation method  used to calculate $\mathbf{y}^{\rm cal} (\mathbf{x})$.
As discussed in Ref.~\citen{Tamura-2016},
the steepest descent method is a promising way for
this calculation when the input data is assumed to be explained by a simple classical Hamiltonian at the zero temperature.
Evaluating $\mathbf{y}^{\rm cal} (\mathbf{x})$ for
the given  $\mathcal{H} (\mathbf{x})$, in general, requires
statistical/quantum mechanical many-body calculations, such as 
the Markov-chain Monte Carlo (MCMC) method\cite{Sandvik-1991,Wang-2001,Kawashima-2004,Suwa-2010,Landau-2014}, 
the exact diagonalization method\cite{Lin-1990,Jaklic-1994,Yamaji-2014}, 
and the density matrix renormalization group method\cite{White-1992,Nishino-1995,Nishino-1996}.
All of which drastically increase the computational cost for the inner loop.

In the outer loop, a sampling of a model parameter $\mathbf{x}$ in the posterior distribution is performed.
In Ref.~\citen{Tamura-2016}, we used the MCMC method with the exchange Monte Carlo method\cite{Hukushima-1996}.
Although this combined method efficiently yields the global maximum of
the probability distribution even when many local maxima exist,
an enormous number of sampling points is very time consuming.
Consequently, the MCMC approach to calculate the outer loop of a complicated effective model estimation is one of the main obstacles for applications in material science.


In this paper, 
a computational method that estimates the effective model with a reduced
outer loop computational cost is discussed on the basis of 
a Bayesian optimization for computationally extensive probability distributions.
In the our Bayesian optimization technique, 
extreme values of acquisition functions obtained by Gaussian processes are used as candidates of maximizers of Eq.~(\ref{eq:cond_model}) or
minimizers of Eq.~(\ref{eq:energy_function}). 
We investigate the efficiency of our Bayesian optimization technique to
search the minimizer of $E (\mathbf{x})$ defined by
Eq.~(\ref{eq:energy_function}) relative to the random search method, the
steepest descent method, and the Monte Carlo method when the number of
sampling points is fixed to be small.
In our demonstrations, the magnetization curve from the classical Ising model calculated by the mean-field approximation and the specific heat from the quantum Heisenberg model calculated by the exact diagonalization method are treated as the inputted measured data.
Consequently,
it is found that the Bayesian optimization is useful to search a better maximizer of the computationally extensive probability distribution.


\section*{Bayesian optimization}


\subsection*{Gaussian process}

The Gaussian processes are a powerful machine learning technique to estimate unknown data from known data sets\cite{Bishop-2006}.
Here we consider the case when the given data set is $\{ \mathbf{x}_n, E (\mathbf{x}_n) \}_{n=1, ..., N}$, where $N$ is the number of data.
In our case,
$\mathbf{x}_n$ is the set of model parameters in the effective physical model and $E (\mathbf{x}_n)$ denotes the value of the energy function $E (\mathbf{x})$ defined by Eq.~(\ref{eq:energy_function}) on $\mathbf{x}_n$. 
Using Gaussian processes which are zero mean,
the conditional probability  $P(E(\mathbf{x}) | \mathbf{x})$ of $E(\mathbf{x})$ given any $\mathbf{x}$ is written as the Gaussian distribution with a mean of $\mu (\mathbf{x})$ and a standard deviation of $\delta (\mathbf{x})$:
\begin{eqnarray}
\mu (\mathbf{x}) &=& \mathbf{k}^{\rm T} (\mathbf{K} + \lambda \mathbf{I}_N)^{-1} \mathbf{E}, \\
\delta^2 (\mathbf{x}) &=& c - \mathbf{k}^{\rm T} (\mathbf{K} + \lambda \mathbf{I}_N)^{-1} \mathbf{k},
\end{eqnarray}
where $\mathbf{I}_N$ is the $N$-dimensional identity matrix.
Furthermore, $\mathbf{E}$, $\mathbf{k}$, $\mathbf{K}$, and $c$ are defined as
\begin{eqnarray}
\mathbf{E} &=& \begin{pmatrix}  E(\mathbf{x}_1) & \cdots & E(\mathbf{x}_N) \end{pmatrix}^{\rm T}, \\
\mathbf{k} &=& \begin{pmatrix} k (\mathbf{x}_1, \mathbf{x}) & \cdots & k (\mathbf{x}_N, \mathbf{x}) \end{pmatrix}^{\rm T}, \\
\mathbf{K} &=& \begin{pmatrix} 
k (\mathbf{x}_1, \mathbf{x}_1) & \cdots & k (\mathbf{x}_1, \mathbf{x}_N) \\
\vdots & \ddots & \vdots \\
k (\mathbf{x}_N, \mathbf{x}_1) & \cdots & k (\mathbf{x}_N, \mathbf{x}_N)
\end{pmatrix}, \\
c &=& k (\mathbf{x},\mathbf{x}) + \lambda,
\end{eqnarray}
where $k (\mathbf{x}_i, \mathbf{x}_j)$ is the Gauss kernel function:
\begin{eqnarray}
k (\mathbf{x}_i, \mathbf{x}_j) &=& \exp \left[ - \frac{1}{2 \gamma^2} \| \mathbf{x}_i - \mathbf{x}_j \|^2 \right].
\end{eqnarray}
Although the computational cost of Gaussian processes is $\mathcal{O} (N^3)$,
some methods to reduce it including an approximation method and their
efficiencies are currently under
investigation~\cite{Rahimi-2007,Heaton-2017}.
In this formula,
$\lambda$ and $\gamma$ are the hyperparameters, which should be specified prior to the analysis.
While various methods have been proposed to determine the
hyperparameters, we adopt the cross validation method for determination of the hyperparameters $\lambda$ and $\gamma$,
which are chosen so as to minimize the prediction error.
In the cross validation,
the data set $D$, that is, $\{ \mathbf{x}_n, E (\mathbf{x}_n) \}_{n=1, ..., N}$ is randomly divided into $S$ data subsets.
Each data subset is expressed by $D_s$ labeled by $s=1, ...,S$. 
One of the $S$ data subsets is regarded as the testing data,
while the remaining $S-1$ subsets are used as training data. 
For each data subset $G_s=D\setminus D_s$,
Gaussian process training is performed when the training data are $\{ \mathbf{x}_n, E (\mathbf{x}_n) \} _{n \in G_s}$.
The mean-square error between the testing data $E (\mathbf{x}_n)$ and the estimated $\mu (\mathbf{x}_n)$ for $n \in D_s$ is evaluated.
The cross validation regards  
the mean-square error as the prediction error when the testing data $D_s$ is treated as unknown data.
The optimal values of $\lambda$ and $\gamma$ are evaluated to minimize
the prediction error averaged over $S$ data subsets.


\subsection*{Bayesian optimization for computationally extensive probability distributions}

We introduce a Bayesian optimization technique to find a better minimizer for the energy function $E (\mathbf{x})$ defined by Eq.~(\ref{eq:energy_function}), 
when the number of sampling points is limited.
Our Bayesian optimization is comprised of the following procedure:

\begin{quote}
\noindent
\textbf{Step 1}:
Sets of model parameters $\mathbf{x}_n$ are randomly generated with
$n=1, ..., P$,
and $E (\mathbf{x}_n)$ is calculated for the generated $\mathbf{x}_n$.
That is, the $P$ calculations of $\mathbf{y}^{\rm cal} (\mathbf{x}_n)$ from $\mathcal{H} (\mathbf{x}_n)$ are necessary.

\noindent
\textbf{Step 2}:
Gaussian process is trained for the data set $\{ \mathbf{x}_n, E (\mathbf{x}_n) \}_{n=1, ..., P}$, 
yielding the mean value $\mu (\mathbf{x})$ and the standard deviation
$\delta (\mathbf{x})$ of $P(E (\mathbf{x}) | \mathbf{x})$.

\noindent
\textbf{Step 3}:
The steepest descent method with randomly chosen initial parameters is
performed for the three types of acquisition functions~\cite{Lai-1985,Benassi-2011,Srinivas-2012,Shahriari-2016} defined as
\begin{eqnarray}
f_{\rm LCB} (\mathbf{x}) &=& \mu (\mathbf{x}) - \kappa \delta (\mathbf{x}), \label{eq:def_fx} \\
f_{\rm GP-LCB} (\mathbf{x}) &=& \mu (\mathbf{x}) - \kappa_t \delta (\mathbf{x}), 
\ \ \ \kappa_t = \sqrt{2 \log (|X| t^2 \pi^2 / 6 \epsilon)}, \label{eq:def_fx2} \\
f_{\rm EI} (\mathbf{x}) &=& - \delta (\mathbf{x}) [ Z \Phi (Z) - \phi (Z) ],
\ \ \ Z= [E_{\rm min} - \mu (\mathbf{x})]/\delta (\mathbf{x}), \label{eq:def_fx3} \ \ \ 
\end{eqnarray}
where $\kappa >0$ and $0 < \epsilon \le 1$ are the hyperparameters.
$|X|$ is the size of the search space, and $t$ is the step of repetition of BO.
Furthermore,
$\phi (Z)$ and $\Phi (Z)$ are the standard normal probability distribution function and its cumulative distribution function, respectively,
and $E_{\rm min}$ is the present minimum value of $E (\mathbf{x})$.
Then, a local or global minimum $\mathbf{x}^*$ of acquisition functions is obtained
and $Q$ different model parameters are generated by repeating this operation.
Note that the fixed value of $\epsilon$ as 0.5 is used in
the analysis of this paper for simplicity.

\noindent
\textbf{Step 4}:
$E (\mathbf{x}^*)$ is calculated for each $\mathbf{x}^*$ obtained in Step 3.
By adding the new data,
the data set is updated as $\{ \mathbf{x}_n, E (\mathbf{x}_n) \}_{n=1, ..., P+Q}$.
Here, the $Q$ calculations of $\mathbf{y}^{\rm cal} (\mathbf{x}_n)$ from $\mathcal{H} (\mathbf{x}_n)$ are necessary.

\noindent
\textbf{Step 5}:
Steps 2--4 are repeated $R$ times.
In each iteration,
the number of data points is increased by $Q$ evaluation.

\noindent
\textbf{Step 6}:
Finally, the minimum value of $E (\mathbf{x})$ from $\{ \mathbf{x}_n, E (\mathbf{x}_n) \}_{n=1, ..., P + Q \times R}$ is determined.

\end{quote}

\noindent
We emphasize that the number of calculations of $\mathbf{y}^{\rm cal} (\mathbf{x}_n)$ from $\mathcal{H} (\mathbf{x}_n)$ is $N_{\rm s}=P + Q \times R$ in this procedure, 
which corresponds to the number of sampling points on $E (\mathbf{x})$.
The computational cost in Step 3 is low because $\mu (\mathbf{x})$ and $\delta (\mathbf{x})$ are quickly obtained for a given $\mathbf{x}$.
Thus, 
many candidates for a local minimum or a global minimum of $E (\mathbf{x})$ are generated
from the acquisition functions without calculation of $E (\mathbf{x})$, 
which is the key of our Bayesian optimization.
Notice that an alternative
approach has been proposed for optimizing a continuous function with an
easily-calculable statistical function
defined only on discrete grid points, in contrast to the our method~\cite{Cox-1992}.


\section*{Results}


\subsection*{Application for posterior distribution based on a classical Ising model}

\begin{figure}[!h]
\includegraphics[width=12cm]{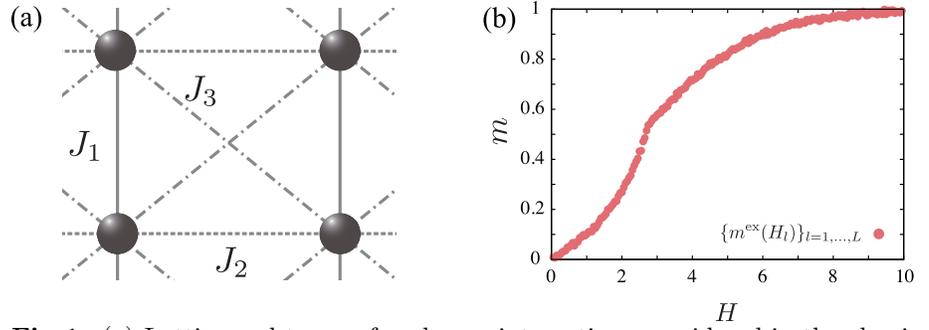}
\caption{
(a) Lattice and types of exchange interactions considered in the classical Ising model defined by Eq.~(\ref{eq:ham_cl}).
(b) Inputted magnetization curve $\{ m^{\rm ex} (H_l) \}_{l=1, ...,L}$
 with $L=200$ where $( x_1, x_2, x_3 ) = ( -1.0, -0.5, 0.3 )$ are used for a temperature $T=3.0$.
}
\label{fig:classical_lattice}
\end{figure}

We demonstrate an application for posterior distribution in effective
physical model estimation based on a classical Ising model in two dimensions.
The model Hamiltonian of the classical Ising model under magnetic field $H$ is defined by
\begin{eqnarray}
\mathcal{H}_{\rm C} (\mathbf{x}) = - \sum_{i,j} J_{ij} \sigma_i^z \sigma_j^z - H \sum_{i} \sigma_i^z, \ \ \ (\sigma_i^z = \pm 1), \label{eq:ham_cl}
\end{eqnarray}
where $J_{ij}$ is the exchange interactions between the $i$-th spin and the $j$-th spin.
Here, we consider three types of exchange interactions
on the square lattice shown in Fig~\ref{fig:classical_lattice} (a). 
In this case,
three different model parameters are to be estimated, that is,
$\mathbf{x}= (x_1, x_2, x_3 ) = (J_1, J_2 , J_3)$.

To discuss the efficiency of the proposed method for 
the effective model estimation,
a synthesis magnetization curve $\{ m^{\rm ex} (H_l) \}_{l=1, ..., L}$ is used as the input data generated by the same model of Eq.~(\ref{eq:ham_cl}).
By performing mean-field calculations for the four sublattice model,
the magnetic field dependency of the magnetization is calculated 
with $( x_1, x_2, x_3 ) = ( -1.0, -0.5, 0.3 )$ for a temperature $T=3.0$.
Here, the Boltzmann constant is set to unity and the physical energy
unit is set to $|J_1|$. 
Gaussian noise with a mean of zero and a standard deviation of 0.004 is added to the obtained magnetization curve.
Fig~\ref{fig:classical_lattice} (b) shows the inputted magnetization curve $\{ m^{\rm ex} (H_l) \}_{l=1, ..., L}$ where the number of data points is $L=200$.

To estimate the effective model from $\{ m^{\rm ex} (H_l) \}_{l=1, ..., L}$, 
we search the maximizer of the posterior distribution, which is defined as
\begin{eqnarray}
P(\mathbf{x} | \{ m^{\rm ex} (H_l) \}_{l=1, ..., L}) &\propto& \exp \left[ - E_{\rm C} (\mathbf{x}) \right], \\
E_{\rm C} (\mathbf{x}) &=& \frac{1}{2 \sigma^2} \sum_{l=1}^L \left( m^{\rm ex}(H_l) - m^{\rm cal} (H_l, \mathbf{x}) \right)^2 - \log P (\mathbf{x}),
\end{eqnarray}
where  $\{ m^{\rm cal} (H_l, \mathbf{x}) \}_{l=1, ...,L}$ is the set of calculated magnetization curves from $\mathcal{H}_{\rm C} (\mathbf{x})$.
In this demonstration, the mean-field calculations for the four sublattice model are used as the inner loop calculation method to obtain $\{ m^{\rm cal} (H_l, \mathbf{x}) \}_{l=1, ...,L}$.
Furthermore, instead of treating the posterior distribution itself,
the minimizer of $E_{\rm C} (\mathbf{x})$ is searched.
For simplicity, the prior distribution of model parameters $P(\mathbf{x})$ is assumed to be a uniform distribution; that is, $P(\mathbf{x}) = 1$
which corresponds to the least square fitting, 
and then the factor $1/2 \sigma^2$ is set to be a constant without loss of generality.

The minimum values of $E_{\rm C} (\mathbf{x})$ obtained by the random
search method, the steepest descent method, the Monte Carlo method, and
the our Bayesian optimization are compared, depending on the number of
sampling points $N_{\rm s}$ on $E_{\rm C} (\mathbf{x})$.
The details of each method are denoted below.

\begin{quote}

\noindent
\textbf{Random search method}: 
A set of model parameters $\mathbf{x}_n = (x_1, x_2, x_3)$ is randomly generated from the region where $-5 \le x_1, x_2, x_3 \le 5$.
Then $E_{\rm C} (\mathbf{x}_n)$ is calculated.
This procedure is repeated $N_{\rm s}$ times, and the data set $\{ \mathbf{x}_n, E_{\rm C} (\mathbf{x}_n) \}_{n=1, ..., N_{\rm s}}$ is obtained, 
from which the minimum value of $E_{\rm C} (\mathbf{x})$ is searched.

\noindent
\textbf{Steepest descent method}: 
An initial set of model parameters [i.e. $\mathbf{x}_1 = (x_1, x_2, x_3)$] is randomly generated from the region where $-5 \le x_1, x_2, x_3 \le 5$. 
A set of model parameters is updated $N_{\rm s}/2$ times by using the following equation from $\mathbf{x}_n = (x_1, ..., x_k, ..., x_K)$ to $\mathbf{x}_{n+1} = (x_1, ..., x_k', ..., x_K)$:
\begin{eqnarray}
x_k' &=& x_k - \alpha \frac{\Delta E}{\Delta x}, \\
\Delta E &=& E_{\rm C} (x_1, ..., x_k+ \Delta x, ..., x_K) - E_{\rm C} (x_1, ..., x_k, ..., x_K).
\end{eqnarray}
Here, $k$ is randomly chosen from $k \in 1, ...,K$ where $K=3$ in this case, and $\Delta x = \alpha = 0.01$.
Notice that the calculation of $E_{\rm C} (\mathbf{x})$ should be repeated twice in each update.
Thus, when the number of updates is $N_{\rm s}/2$, the number of sampling points on $E_{\rm C} (\mathbf{x})$ becomes $N_{\rm s}$.
Using this update of the model parameters, $E_{\rm C} (\mathbf{x})$ decreases for each update.
From the obtained $\{ \mathbf{x}_n, E_{\rm C} (\mathbf{x}_n) \}_{n=1, ..., N_{\rm s}/2}$, the minimum value of $E_{\rm C} (\mathbf{x})$ is searched.

\noindent
\textbf{Monte Carlo method}: 
An initial set of model parameters [i.e. $\mathbf{x}_1 = (x_1, x_2, x_3)$] is randomly generated from the region where $-5 \le x_1, x_2, x_3 \le 5$. 
A set of model parameters is updated $N_{\rm s}$ times using the following Metropolis-type transition probability from $\mathbf{x}_n$ to $\mathbf{x}_{n+1}$:
\begin{eqnarray}
w (\mathbf{x}_{n+1} | \mathbf{x}_{n}) &=& \min \left\{ 1, \exp \left[ - \Delta E(\mathbf{x}_{n+1},\mathbf{x}_{n}) \right] \right\}, \\
\Delta E(\mathbf{x}_{n+1},\mathbf{x}_{n}) &=& E_{\rm C} (\mathbf{x}_{n+1})-E_{\rm C} (\mathbf{x}_{n}).
\end{eqnarray}
Here, the set of model parameters after updating is prepared as $\mathbf{x}_{n+1} = (x_1, ..., x_k', ..., x_K)$ with $x_k' = x_k + r$ from the set of model parameters before updating $\mathbf{x}_n = (x_1, ..., x_k, ..., x_K)$, 
where $k$ is randomly chosen from $k \in 1, ...,K$, and $r$ is a random number between $-1$ and $+1$.
From the obtained $\{ \mathbf{x}_n, E_{\rm C} (\mathbf{x}_n) \}_{n=1, ..., N_{\rm s}}$, the minimum value of $E_{\rm C} (\mathbf{x})$ is searched.

\noindent
\textbf{Bayesian optimization}: 
A set of model parameters $\mathbf{x}_n$ is randomly generated from the region where $-5 \le x_1, x_2, x_3 \le 5$ and $E_{\rm C} (\mathbf{x}_n)$ is calculated.
This procedure is repeated $P=200$ times as the initial data set, 
and the Bayesian optimization is performed with $Q=10$ and $R=(N_{\rm s}-P)/Q$.
In the method, the steepest descent method in Step 3 is implemented by using the following equation from $\mathbf{x} = (x_1, ..., x_k, ..., x_K)$ to $\mathbf{x}' = (x_1, ..., x_k', ..., x_K)$:
\begin{eqnarray}
x_k' &=& x_k - \alpha \frac{\Delta f}{\Delta x}, \\
\Delta f &=& f (x_1, ..., x_k+\Delta x, ..., x_K) - f (x_1, ..., x_k, ..., x_K),
\end{eqnarray}
where $f(\mathbf{x})$ expresses the acquisition functions defined by Eqs.~(\ref{eq:def_fx}), (\ref{eq:def_fx2}), and (\ref{eq:def_fx3}).
Here, $k$ is randomly chosen from $k \in 1, ...,K$, and $\Delta x = \alpha = 0.01$.
$f (\mathbf{x})$ is defined by Eq.~(\ref{eq:def_fx}), which is obtained from Gaussian process.
In our calculation, the steepest descent method is performed with 100 updates to obtain the extreme value of $f (\mathbf{x})$.
From the obtained $\{ \mathbf{x}_n, E_{\rm C} (\mathbf{x}_n) \}_{n=1, ..., N_{\rm s}}$, the minimum value of $E_{\rm C} (\mathbf{x})$ is searched.

\end{quote}

\begin{figure}[!h]
\includegraphics[width=12cm]{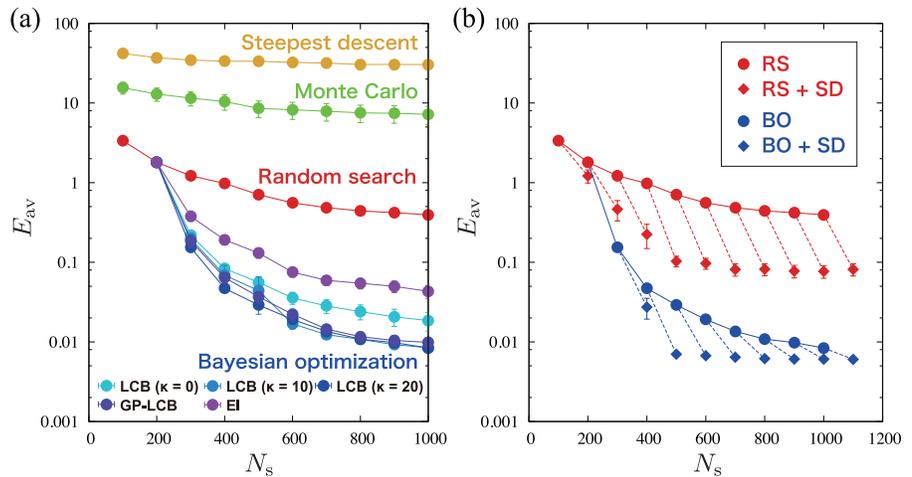}
\caption{
Results of the average $E_{\rm av}$ of the minimum values of $E_{\rm C} (\mathbf{x})$ obtained from 100 independent runs in the effective model estimation of the classical Ising model. 
(a) $E_{\rm av}$ as a function of $N_{\rm s}$,
which is the number of sampling points on $E_{\rm C} (\mathbf{x})$,  obtained from the random search method (red circles), the steepest descent method (yellow circles), the Monte Carlo method (green circles), and the Bayesian optimization (blue circles).
(b) $E_{\rm av}$ as a function of $N_{\rm s}$ obtained from the random search method (RS) (red circles), the Bayesian optimization using $f_{\rm LCB} (\mathbf{x})$ with $\kappa=20$ (BO) (blue circles), the random search method with the steepest descent method (RS$+$SD) (red diamonds), and the Bayesian optimization with the steepest descent method (BO$+$SD) (Blue diamonds).
Dashed lines connect the initial $E_{\rm av}$ (circle point) by only RS or BO and the obtained $E_{\rm av}$ (diamond point) by performing the steepest descent method with 50 updates after RS or BO.
}
\label{fig:classical_eav}
\end{figure}

Fig~\ref{fig:classical_eav} (a) is the sampling number $N_s$
dependence of the averaged minimum value $E_{\rm av}$ of $E_{\rm C} (\mathbf{x})$ for 100 independent runs with each methods.
The error bars are calculated from the standard deviation.
The Bayesian optimization yields the smallest $E_{\rm av}$,
indicating that the Bayesian optimization gives better minimizers of $E_{\rm C} (\mathbf{x})$ even if $N_s$ is small.
Furthermore, the most successful analysis is given by the Bayesian optimization using $f(\mathbf{x})_{\rm LCB}$ with $\kappa=20$, 
while the steepest descent method and the Monte Carlo method produce worse results than the random search method.
These methods are frequently trapped at a local minimum
depending on the initial set of model parameters,
and eventually $E_{\rm av}$ stays at large values.

\begin{figure}[!h]
\includegraphics[width=12cm]{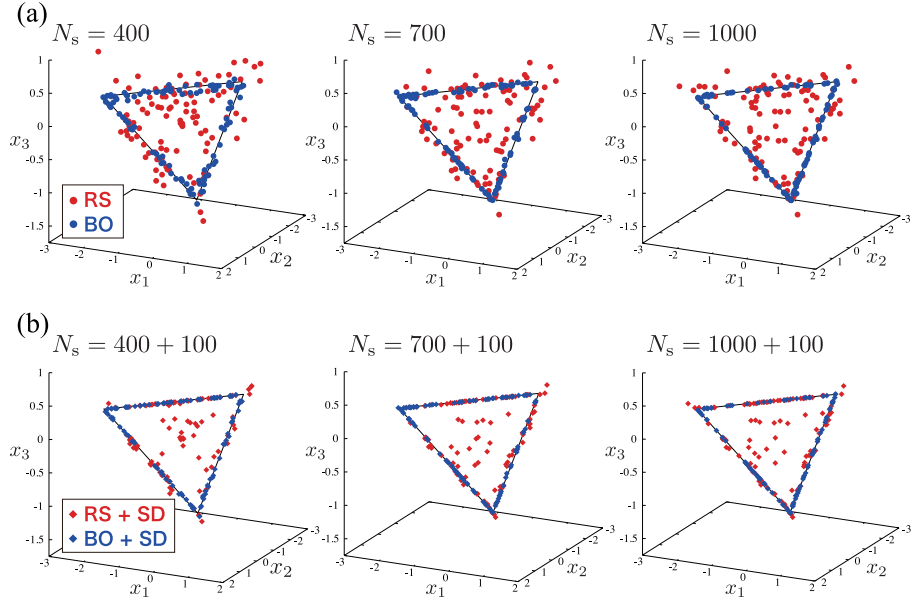}
\caption{
Results of the estimated model parameters in the effective model estimation based on the classical Ising model.
(a) Distribution of the estimated model parameters from 100 independent runs depending on $N_{\rm s}$ by the random search method (RS) (red circles) and the Bayesian optimization using $f_{\rm LCB} (\mathbf{x})$ with $\kappa=20$ (BO) (blue circles).
The black lines indicate exact solutions when the input magnetization curve without Gaussian noise is obtained.
(b) Distribution of the estimated model parameters by the random search method with the steepest descent method (RS$+$SD) (red diamonds) and the Bayesian optimization with the steepest descent method (BO$+$SD) (blue diamonds).
In these cases,
starting from the results shown in (a) by RS and BO, 
the steepest descent method is further performed with 50 updates. 
}
\label{fig:classical_interactions}
\end{figure}

Fig~\ref{fig:classical_interactions} (a) is the distribution of the estimated model parameters for 100 independent runs with various
$N_{\rm s}$ by the random search method and the Bayesian optimization.
The black lines indicate exact solutions by which the input magnetization curve without Gaussian noise is
generated, 
except for the case where any one of the parameters $x_k$ has zero. 
As $N_{\rm s}$ increases,
the results by the Bayesian optimization converge on the black lines, implying that the model parameters can be correctly estimated with a high probability.
On the other hand, the case of the random search method shows no
significant improvement with increasing $N_{\rm s}$.
This could be understood by noticing that the accuracy of the acquisition functions by Gaussian processes in the
Bayesian optimization is
improved with increasing the sampling points, namely $N_{\rm s}$, while
the random search method does not refer to the prior sampling points.

The Bayesian optimization  as well as the random search method, in
general, does not take into account local structure of the energy function such
as gradient in the parameter space. 
To improve the solutions, we consider combinations of the steepest descent method with the random search method or the Bayesian optimization.
One may expect that the steepest descent method produces a local minimum or a global minimum around the estimated model parameters by the random search method or the Bayesian optimization.
That is, the estimated model parameters by the random search method or
the Bayesian optimization are used as the initial set of model
parameters in the steepest descent method, which  is performed with 50 updates.
Fig~\ref{fig:classical_eav} (b) compares $E_{\rm av}$'s by the random search method, the Bayesian optimization using $f_{\rm LCB} (\mathbf{x})$ with $\kappa=20$, 
and those with the steepest descent method.
The drastic improvement can be confirmed even for 50 updates in the steepest descent method.
Note that if the number of updates in the steepest descent method is increased, the obtained $E_{\rm av}$ should be improved. 
However, since the number of sampling is also increased, 
a trade-off between search for initial sets by the random search method or the Bayesian optimization and evaluation of local structures by the steepest descent method should be optimized. 
We confirmed for some cases that the Bayesian optimization with steepest descent method is the best among the considered methods.

Fig~\ref{fig:classical_interactions} (b) shows the distribution of the estimated model parameters.
For the Bayesian optimization with the steepest descent method, 
the real minimizer of $E_{\rm C} (\mathbf{x})$ is found in all independent runs,  
while some of the obtained results by the random search method with the steepest descent method differ from the exact solutions, 
and these cases are trapped in local minima.
The steepest descent method significantly improves the estimates by the
Bayesian optimization and random search methods. 
The results imply that the Bayesian optimization combined with the steepest descent method is powerful tool to find the global minimum of $E_{\rm C} (\mathbf{x})$.


\subsection*{Application for posterior distribution based on a quantum Heisenberg model}

\begin{figure}[!h]
\includegraphics[width=12cm]{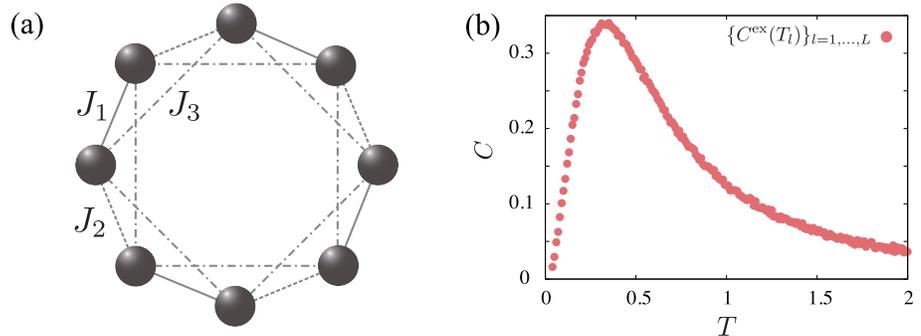}
\caption{
(a) Lattice and types of exchange interactions considered in the quantum Heisenberg model defined by Eq.~(\ref{eq:ham_qu}).
(b) Inputted specific heat result $\{ C^{\rm ex} (T_l) \}_{l=1, ...,L}$
with $L=200$ and
$( x_1, x_2, x_3, x_4, x_5 ) = ( 1.0, 0.8, -0.2, -0.7, 0.3 )$.
}
\label{fig:quantum}
\end{figure}

The case where the number of model parameters increases against the previous case is considered when a quantum Heisenberg model on the one-dimensional chain is used (Fig~\ref{fig:quantum} (a)).
The model Hamiltonian of the quantum Heisenberg model under magnetic field $H$ is defined by
\begin{eqnarray}
\mathcal{H}_{\rm Q} (\mathbf{x}) = - \sum_{i,j} J_{ij} \left[ \hat{\sigma}_i^x \hat{\sigma}_j^x + \hat{\sigma}_i^y \hat{\sigma}_j^y + \Delta \hat{\sigma}_i^z \hat{\sigma}_j^z \right] - H \sum_{i} \hat{\sigma}_i^z, \label{eq:ham_qu}
\end{eqnarray}
where $\Delta$ is the parameter for the anisotropy and $(\hat{\sigma}_i^x, \hat{\sigma}_i^y, \hat{\sigma}_i^z)$ is the Pauli matrix.
Here,
the model parameters are $\mathbf{x}=(x_1, x_2, x_3, x_4, x_5) = ( J_1, J_2, J_3, \Delta, H )$.
Fig~\ref{fig:quantum} (a) depicts three types of exchange interactions.

This demonstration uses the temperature dependence of the specific heat as an input data. 
The input specific heat $\{ C^{\rm ex} (T_l) \}_{l=1, ..., L}$ is generated from the model defined by Eq.~(\ref{eq:ham_qu}) as follows.
By performing the exact diagonalization method,
the temperature dependence of the thermal average of the specific heat 
for $( x_1, x_2, x_3, x_4, x_5 ) = ( 1.0, 0.8, -0.2, -0.7, 0.3 )$ is calculated.
The Gaussian noise with a mean of zero and a standard deviation of 0.004 is added to the obtained specific heat.
Fig~\ref{fig:quantum} (b) shows the temperature dependence of the specific heat with $L=200$, 
which is used as the input in the effective model estimation.
As shown in the previous case, our task is to search for 
the minimizer of energy function $E_{\rm Q} (\mathbf{x})$ defined as
\begin{eqnarray}
E_{\rm Q} (\mathbf{x}) &=& \sum_{l=1}^L \left( C^{\rm ex}(T_l) - C^{\rm cal} (T_l, \mathbf{x}) \right)^2, 
\end{eqnarray}
where 
$\{ C^{\rm cal} (T_l, \mathbf{x}) \}_{l=1, ...,L}$ is the set of calculated specific heat from $\mathcal{H}_{\rm Q} (\mathbf{x})$ by performing the exact diagonalization method.

\begin{figure}[!h]
\includegraphics[width=12cm]{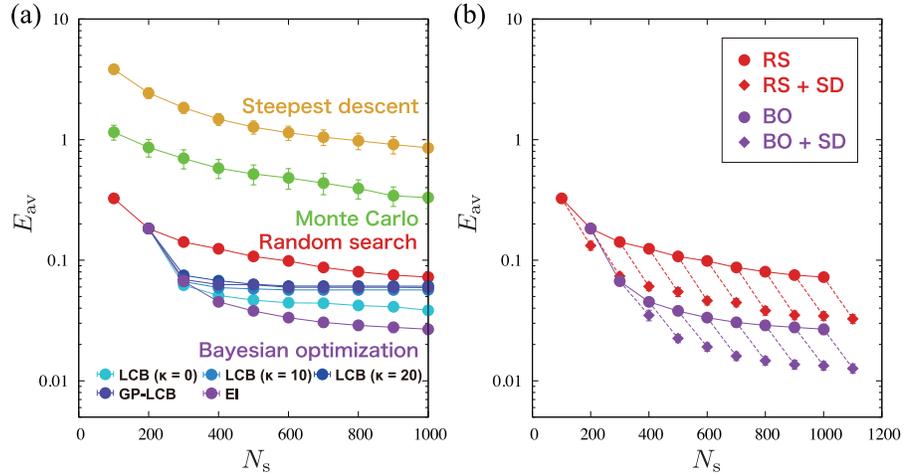}
\caption{
Results of the average $E_{\rm av}$ of the minimum values of $E_{\rm Q} (\mathbf{x})$ obtained from 100 independent runs in the effective model estimation of the quantum Heisenberg model.
(a) $E_{\rm av}$ as a function of $N_{\rm s}$ obtained from the random search method (red circles), the steepest descent method (yellow circles), the Monte Carlo method (green circles), and the Bayesian optimization (blue circles).
(b) $E_{\rm av}$ as a function of $N_{\rm s}$ obtained from the random search method (RS) (red circles), the Bayesian optimization using $f_{\rm EI} (\mathbf{x})$ (BO) (blue circles), the random search method with the steepest descent method (RS$+$SD) (red diamonds), and the Bayesian optimization with the steepest descent method (BO$+$SD) (blue diamonds).
In the steepest descent method, 50 updates are performed after RS or BO.
}
\label{fig:quantum_eav}
\end{figure}

We compared $E_{\rm av}$, which is the average of the minimum value of $E_{\rm Q} (\mathbf{x})$ for 100 independent runs, for the random search method, the steepest descent method, the Monte Carlo method, and the Bayesian optimization (Fig~\ref{fig:quantum_eav} (a)).
The setups of these methods are the same as the previous case except for the number of model parameters ($K=5$) and the region in which a set of model parameters is randomly generated.
In this case, we use $-3 \le x_1, x_2, x_3 \le 3$ and $-2 \le x_4, x_5 \le 2$.
The results are qualitatively the same as the previous case.
The most successful analysis is produced by the Bayesian optimization using $f(\mathbf{x})_{\rm EI}$.
This result is different from the previous demonstration, 
which means that
an appropriate acquisition function depends
on a target physical model and input physical quantities.
Furthermore, as shown in Fig~\ref{fig:quantum_eav} (b),
the combined steepest descent method improves the estimates of the
Bayesian optimization and the random search method again. 
Similar to the previous case,
the Bayesian optimization with the steepest descent method gives a better minimizer of $E_{\rm Q} (\mathbf{x})$.
Consequently, we conclude that the Bayesian optimization is useful to find a better maximizer of the posterior distribution in an effective model estimation with a small number of sampling points.

\section*{Discussion}

We searched for a better maximizer of a posterior distribution in the effective physical model estimation which is a computationally extensive probability distribution, using the Bayesian optimization.
It is found for at least  two simple models that the Bayesian
optimization has a higher efficiency of finding a better
maximizer of the posterior distribution compared to the random search method, 
the steepest descent method, and the Monte Carlo method when the number
of sampling points on the posterior distribution is fixed to be small,
while an appropriate acquisition function providing
a high efficiency still depends on the problem to be solved.
Our Bayesian optimization has some hyperparameters, i.e., $P, Q$, and $R$. 
Although we did not optimize these hyperparameters, 
the Bayesian optimization is a better method to obtain the maximizer of the posterior distribution.
Particularly, since the value of $Q$ is related to the batch/parallel problem of the Bayesian optimization\cite{Chevalier-2012,Desautels-2014},
some improvement of the performance is expected by tuning $Q$.
Furthermore, 
a combination of the Bayesian optimization and the steepest descent
method drastically increases the efficiency of finding a better maximizer of the posterior distribution.
The key of our Bayesian optimization is to predict a set of model parameters near a local maximum or a global
maximum of the posterior distribution from the extreme values of acquisition functions by Gaussian processes, 
which requires a relatively low computational cost. 
Consequently,
the model parameters near a global maximum can be found with a high probability.
These facts suggest that the Bayesian optimization will be a powerful tool for effective model estimations.
However, 
to find a maximizer of posterior distributions with various types of prior distributions and a large number of model parameters, 
the Bayesian optimization may be not always useful.
Then in the future, we will evaluate effective model estimations using the Bayesian optimization for actual materials.
Because the maximizer of a probability distribution is searched in many scientific fields,
the Bayesian optimization will play an important role in the promotion of science.


\section*{Acknowledgments}

We thank Shu Tanaka for the useful discussions.
R. T. was partially supported by the Nippon Sheet Glass Foundation for Materials Science and Engineering. 
K. H. was partially supported by a Grants-in-Aid for Scientific Research
from JSPS, Japan (Grant No. 25120010 and 25610102). 
The computations in the present work were performed on Numerical Materials Simulator at NIMS, 
and the supercomputer at Supercomputer Center, Institute for Solid State Physics, The University of Tokyo.
This work was done as part of the ``Materials Research by Information Integration'' Initiative of the Support Program for Starting Up Innovation Hub, Japan Science and Technology Agency.

%
%
%

\end{document}